# Second-Harmonic Young's Interference in Atom-Thin Heterocrystals

Wontaek Kim,[1] Je Yhoung Ahn,[1] Juseung Oh,[1] Ji Hoon Shim,[1,2] and Sunmin Ryu*[1]

[1]Department of Chemistry, Pohang University of Science and Technology (POSTECH), Pohang, Gyeongbuk 37673, Korea

[2]Division of Advanced Materials Science, Pohang University of Science and Technology (POSTECH), Pohang, Gyeongbuk 37673, Korea

*Correspondence to: sunryu@postech.ac.kr

**Abstract**

**Second-harmonic generation (SHG) is a nonlinear optical process that converts two identical photons into a new one with doubled frequency. Two-dimensional semiconductors represented by transition metal dichalcogenides are highly efficient SHG media because of their excitonic resonances. Using spectral phase interferometry, here we directly show that SHG in hetero-bilayers of $MoS_2$ and $WS_2$ is governed by optical interference between two coherent SH fields that are phase-delayed differently in each material. We also quantified the frequency-dependent phase difference between the two, which agreed with polarization-resolved data and first-principles calculations on complex susceptibility. The second-harmonic analogue of Young's double-slit interference shown in this work demonstrates the potential of custom-designed parametric generation by atom-thick nonlinear optical materials.**

**Introduction**

Two-dimensional (2D) materials have emerged as promising platforms for various photonic applications such as ultrafast photodetectors of gapless graphene,[1] valleytronics of semiconducting transition metal dichalcogenides (TMDs),[2] and single-photon emitters of insulating hBN.[3] Their interaction with light is further diversified and strengthened not only by their wide-varying electronic structures but also their low dimensionality and reduced dielectric screening.[4] The second-order susceptibility responsible for optical second-harmonic generation (SHG) is also greatly enhanced by the strong excitonic resonances in 2D TMDs[5-7] and sufficiently large for 2D GaSe even in non-resonant conditions.[8] SHG is a nonlinear parametric process that doubles the frequency of incoming light. Only allowed in non-centrosymmetric materials,[9] it has been widely used in frequency modulation of lasers,[10] surface scientific investigation,[11] and label-free imaging in biological and medical sciences.[12] Two-dimensional crystals are ideal SHG-materials[13, 14] not only for their strong light-matter interaction[15] and atomic thickness defying the phase-matching requirement but also for their stackability into customized hetero-crystals with high angular precision and material diversity.[16]

The complex nature of the nonlinear susceptibility arising from the light absorption[17] provides another control, the phase delay between the fundamental and SH waves, to manipulate the photonic process. Since the first measurements on GaAs,[17] the phase information from interferometric SHG has been used in determining absolute molecular orientation,[18] probing magnetization reversal,[19] and imaging nonlinear susceptibilities.[20] More recently, interference in SHG intensity was observed in 2H-stacked[7] and artificially stacked[21] 2D crystals. Destructive interference was imaged along the boundaries between two anti-parallel domains of TMDs,[22] which could be further differentiated by spectral phase interferometry.[23] Despite these, however, there is a lack in fundamental understanding of material-dependence of the phase delay in SHG and how it dictates the superposition of SH fields generated in two dissimilar 2D materials in van der Waals (vdW) contact.

In this work, we report interferometric mixing of SHG signals generated in vdW hetero-crystals consisting of single-layer (1L) MoS$_2$ and WS$_2$. We show that the superposition is governed by SHG phase delays characteristic of materials and photon energy, and directly quantify them by spectral phase interferometry. First-principles calculations also reveal the electronic origin of the phase differences. VdW stacks of 2D crystals are an excellent photonic system not only for distinctive electronic structures but also for facile integration into photonic structures including waveguides[24] and cavities.[25]

**Results & Discussion**

**Anomalous SHG behavior of hetero-bilayers.** As model systems (Fig. 1a & Fig. S1), homo-bilayers of MoS$_2$/MoS$_2$ (2L$_{MoMo}$) and hetero-bilayers of MoS$_2$/WS$_2$ (2L$_{MoW}$) were fabricated on fused quartz substrates by the deterministic dry transfer method[26, 27] (see Methods). The stack angle ($\theta_s$: 0 ~ 60°) between two single layers (1Ls) as defined in Fig. 1b could be controlled within one degree during the transfer step using the crystallographic orientations of each layer determined by SHG measurements. The step height determined from the topographic AFM image in Fig. 1a was 1.0 ± 0.2 nm, which indicated that its vdW gap size of 2L$_{MoW}$ was close to that of 2H-type bilayers,[28] and thus the two 1Ls were in good contact. An average gap size obtained for multiple samples of 2L$_{MoW}$ and 2L$_{MoMo}$ was 1.0 ± 0.1 nm (see Fig. S1 for more samples and also Methods for post-stacking treatments). Raman and photoluminescence spectroscopy showed that individual 1Ls were of high quality and the artificial stacking did not induce significant changes (Fig. S2). As schematically shown in Fig. 1b, the frequency-doubling process was induced in the samples by a plane-polarized fundamental beam (frequency $\omega$) focused with a refractive objective lens. $I_\parallel$, SHG signal parallel to the electric field of the fundamental beam ($E_\omega$), was collected using a polarizer from bilayers and their unstacked 1Ls by varying the azimuthal angle ($\theta$) of $E_\omega$ in the basal plane (Fig. 1b). The second-order susceptibility tensor of $D_{3h}^1$ space group which 1L MoS$_2$ and WS$_2$ belong to requires that $I_\parallel$ is proportional to $\cos^2 3\theta$ and reaches a maximum when $E_\omega$ is parallel to armchair directions ($\overrightarrow{AC}$) as marked in Fig. 1b (see Supplementary Section S1) [7]. Indeed, the unstacked 1L areas of 2L$_{MoMo}$ (Fig. 1c) and 2L$_{MoW}$ (Fig. 1d) obeyed the predicted angular relation exhibiting 6-fold symmetry with angular nodes as $\theta$ was varied by rotating the sample. Then, the difference in the angles for maximum intensities, 33º (34º) for 2L$_{MoMo}$ (2L$_{MoW}$), corresponded to $\theta_s$. It is to be noted that two candidates for stack angle ($\theta_s$ and $\theta_s'$) can be found for two given sets of 6-fold lobes as shown in Fig. S3c. As a convention, $\theta_s$ was defined so that the 6 lobes of bilayers were contained within the angular space spanned by $\theta_s$ (Fig. S3c).

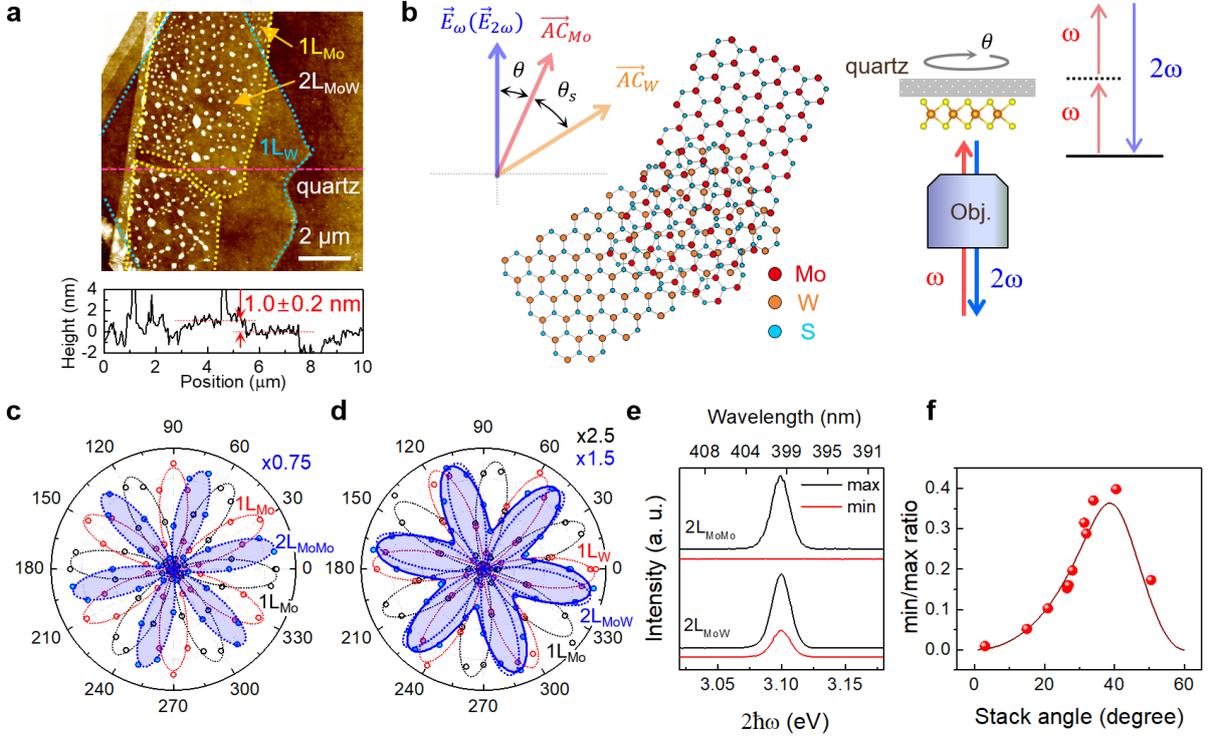

**Fig. 1. Anomalous SHG behavior of hetero-bilayers.** (a) AFM height image of $MoS_2/WS_2$ hetero-bilayer ($2L_{MoW}$) and height profile along the dashed line in the image. The boundaries of $1L_{Mo}$ and $1L_W$ were marked with dotted lines. (b) Schemes for artificially-stacked $2L_{MoW}$ (left) and SHG detection using an objective (right). Samples with stack angle ($\theta_s$) defined against the armchair directions ($\overrightarrow{AC}_{Mo}$ and $\overrightarrow{AC}_W$) of individual layers were rotated to vary azimuthal angle ($\theta$). The polarization of the SH field ($E_{2\omega}$) parallel to that of the fundamental field ($E_\omega$) was selected. (c & d) Polar graphs of SHG signals ($I_\parallel$) from $MoS_2$ homo-bilayer ($2L_{MoMo}$) (c) and $2L_{MoW}$ (d): two individual layers (black and red circles) and bilayers (blue circles). The black and red dotted lines are $\cos^2\theta$-fits to the data. The blue dotted and solid lines represent the superposition of SH fields based on real and complex susceptibilities, respectively (Supplementary Section S2). The blue shades showing agreement between the experiments and the model highlight that the SHG intensity of $2L_{MoW}$ does not reach zero at any angle. (e) SHG spectra with maximum and minimum intensities for $2L_{MoMo}$ and $2L_{MoW}$, respectively. (f) Minimum/maximum ratio in $I_\parallel$ of $2L_{MoW}$ given as a function of $\theta_s$. The solid line represents a theoretical prediction for $\phi_{MoW} = 61°$, which matched with the data best (see Supplementary Section S2). The fundamental wavelength was 800 nm for (c ~ f).

We found that the SHG response of hetero-bilayers was distinctive from that of homo-bilayers. The polar graph of 2L$_{MoMo}$ blue-shaded in Fig. 1c also exhibited 6-fold symmetry with obvious nodes like those of 1Ls. As the bottom and top MoS$_2$ 1Ls are coherently polarized by the fundamental pulse at 800 nm, the SHG signal is the superposition of the SH fields generated in both layers[7, 21]. This interpretation was directly confirmed by the fact that the data of 2L$_{MoMo}$ matched well with the blue dotted line representing the vectorial superposition of the SH fields from the two individual MoS$_2$ 1Ls (see Supplementary Section S2). In contrast, 2L$_{MoW}$ (blue-shaded in Fig. 1d) lacked nodes despite its 6-fold symmetry, which could not be explained by the simple superposition (blue dotted line in Fig. 1d). Notably, its minimum intensity was substantially high (37% of the maximum) unlike that of 2L$_{MoMo}$ which remained typically below 0.5% (Fig. 1e). The anomaly was observed in multiple hetero-bilayers with various stack angles. Whereas all the samples exhibited the 6-fold symmetry (Fig. S3), the minimum/maximum intensity ratio (R) was higher for larger $\theta_s$, but the opposite for $\theta_s > 40º$ as shown in Fig. 1f.

**Elliptical polarization induced by material-dependent SHG phase.** The anomaly suggests that the SH light from hetero-bilayers contains complexity beyond a simple plane polarization. SHG polar graphs remained unchanged (Fig. S4) after vacuum annealing which drastically affected interface quality (Fig. S1). This fact implied that the anomaly is induced by neither charge nor energy transfer. To anatomize the polarization state of SHG signals, we performed polarization-resolved measurements by rotating the analyzing polarizer located in front of the detector (see Methods). As shown in Fig. 2a, $2L_{MoMo}$ obeying the Malus law generated plane-polarized signals like 1Ls, which is consistent with the tensor model (Supplementary Section S1). The signals of $2L_{MoW}$, however, were elliptically polarized with a ratio of $0.37 \pm 0.03:1$ between the minor and major axes irrespective of the sample orientation ($\theta$). This observation is reminiscent of polarization mixing by a quarter-wave plate made of birefringent materials. As depicted in Fig. 2b, two plane-polarized light waves with zero phase difference generate another plane-polarized light. With finite phase difference, however, the superposition leads to a light wave of elliptical polarization in general. Then it can be seen that the phase difference ($\phi_{MoW} = \phi_{Mo} - \phi_W$) between MoS$_2$ and WS$_2$ governs the SHG interference along with the stack angle, as illustrated in Fig. 2c. Note that $\phi_{Mo}$ and $\phi_W$ represent the phase delay of SH fields generated respectively in MoS$_2$ and WS$_2$ with respect to the fundamental fields. Furthermore, one can determine the phase difference from the polarization-resolved data shown in Fig. 1d and Fig. S3 using the interference model of SH waves (Supplementary Section S2). The minimum/maximum intensity ratios (R) in Fig. 1f were best described by the solid line representing $\phi_{MoW} = 61°$ at 800 nm. The value also agreed well with the average (61.0 ± 7.5°) obtained by fitting the data obtained from multiple samples (Fig. S3). This finding reveals that the phase delay between the fundamental and SH waves is substantially dependent on materials. As will be described below, the phase difference also exhibited a strong dependence on photon energy.

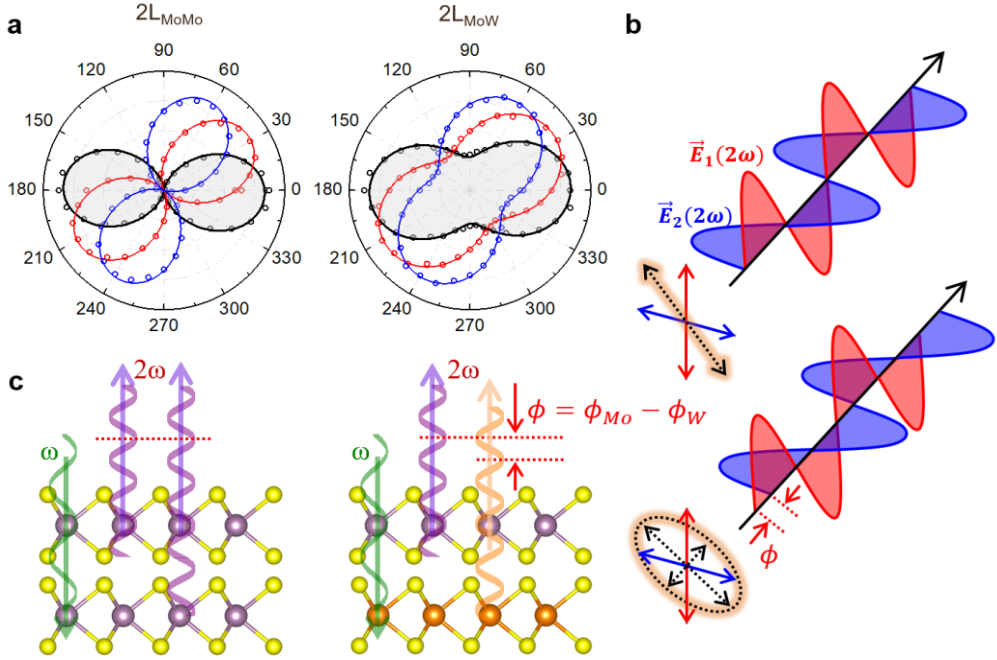

**Fig. 2. Elliptical polarization induced by material-dependent SHG phase.** (a) Polarization analysis of SHG signals from $2L_{MoMo}$ and $2L_{MoW}$. The sample was rotated in a step of 10° to give three different angles. The polar graphs are given as a function of angle ($\theta$) between the analyzing polarizer and the polarization of the fundamental beam. Solid lines are fits to the data: plane polarization ($\cos^2\theta$) for $2L_{MoMo}$ and elliptical polarization ($\cos^2\theta + \varepsilon^2 \sin^2\theta$) for $2L_{MoW}$. (b) Superposition of two plane-polarized SH fields without ($2L_{MoMo}$) and with ($2L_{MoW}$) phase difference. (c) Schematic representations of phase-delayed SHG in $2L_{MoMo}$ (left) and $2L_{MoW}$ (right), respectively.

**Interferometric determination of SHG phase.** Using spectral phase interferometry[17, 19, 23] as an independent and the most definitive probe, we directly measured the phase delay of individual TMD layers of the heterostructures (Supplementary Section S3). As shown in Fig. 3a (see Methods), the reference SHG pulse ($2\omega_{ref}$) generated in an α-quartz crystal was delayed by $\tau$ (2.86 ps for Fig. 3b) behind the sample SHG pulse ($2\omega_{sample}$) because of finite optical dispersion between $\omega$ and $2\omega$ induced by the optical materials shown in Fig. 3a. During diffraction by a grating in the spectrometer, the two coherent pulses with a temporal width of ~100 fs were stretched to ~300 ps and overlapped each other in space and time at the CCD detector plane (Supplementary Section S3). Unlike conventional intensity spectra, the SHG interferograms contained prominent oscillations, as shown in Fig. 3b for $1L_{Mo}$. Whereas the oscillation period of the interferograms in the frequency domain is inversely proportional to $\tau$,[19] the positions of crests and valleys depend on the phase delay defined with respect to the reference SHG signal from α-quartz (Supplementary Section S3). The interferograms in Fig. 3c present the oscillating components only with the rest removed using the Fourier transform analysis. We first confirmed that the interferograms shifted by half of one period when the $1L_{Mo}$ sample was rotated by 60º or its multiples in Fig. 3c (top) (also see Fig. S5a for the phase inversion near 30º). Because such rotations inverse the lattice with respect to the polarization of the fundamental beam, the observation validates that the offset in energy corresponds to the phase difference between the two SHG signals. The phase values of $1L_{Mo}$ were also consistent within three degrees at 800 nm across the sample (Fig. S5b). Even homo-bilayer area of $2L_{MoMo}$ ($\theta_s$ =33°) gave phase values identical to that of each monolayer (Fig. S5c).

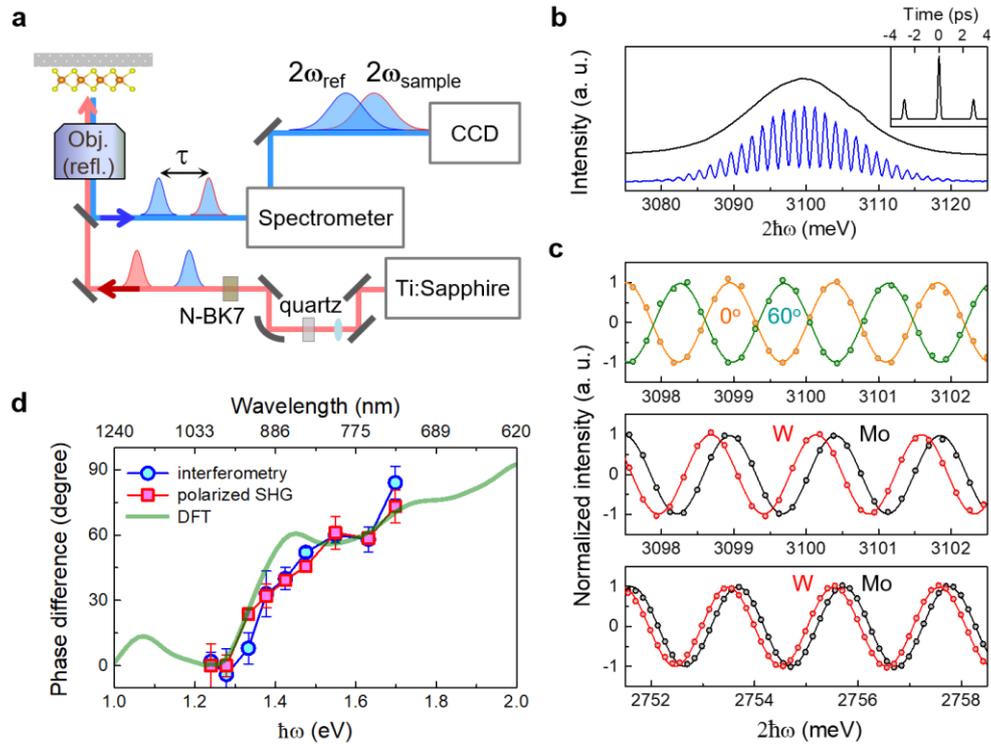

**Fig. 3. Interferometric determination of SHG phase.** (a) Instrumental scheme for spectral phase interferometry (see Methods for details). (b) SHG spectral interferogram (blue) in comparison with a conventional intensity spectrum (black) of $1L_{Mo}$ ($\hbar\omega$ = 1.55 eV). The inset presents the Fourier transform of the interferogram. (c) Interferograms of $1L_{Mo}$ oriented for two inequivalent $\overrightarrow{AC}$ directions through 60º rotation (top). Interferograms of $1L_{Mo}$ (black) and $1L_W$ (red) areas in $2L_{MoW}$ ($\theta_s$ =26.5º): $\hbar\omega$ = 1.55 (middle) and 1.38 eV (bottom). Solid lines are fits to the data (see Supplementary Section S3). (d) SHG phase difference ($\phi_{MoW}$) between $1L_{Mo}$ and $1L_W$ determined independently by the interferometry, polarized SHG measurements (Fig. 1), and first-principles calculations.

Remarkably, the interferograms of $1L_{Mo}$ and $1L_W$ areas in the $2L_{MoW}$ sample showed a substantial offset in energy, as shown in Fig. 3c (middle) (see Fig. S6 for the optical micrograph and raw interferograms). The inter-material phase difference ($\phi_{MoW}$) corresponding to the displacement was 61° at 800 nm and decreased significantly to 32° at 900 nm (Fig. 3c, bottom). In Fig. 3d, we presented two sets of $\phi_{MoW}$ values independently obtained from the interferometry (Fig. 3) and polarized SHG measurements (Fig. 1) for a wide range of fundamental photon energy ($\hbar\omega$). Most of all, both methods yielded highly consistent results for $\phi_{MoW}$, substantiating the interference model involving complex susceptibility (Fig. 2 and Supplementary Section S2). The agreement indicates that the interlayer interactions hardly affect $\phi_{MoW}$ because the interferometry probed 1L regions unlike the angle-resolved SHG. It is also notable that the phase difference drastically decreased and finally reached zero as the photon energy was lowered from 1.55 eV (800 nm) to 1.24 eV (1000 nm). For the highest energy (1.70 eV for 730 nm) that could be handled with the setup, the phase was even larger than that for 1.55 eV.

**Theoretical prediction of SHG phase difference.** To unravel the origin of the energy and material-dependence, we performed first-principles calculations of the second-order susceptibility $\chi^{(2)}$ (see Methods and Supplementary Section S4 for the details of density functional theory calculation). The SHG phase values of both monolayers extracted from the real and imaginary parts of $\chi^{(2)}$ (Fig. S7) remained near zero for energy below 0.7 eV and exhibited a noticeable difference from each other for the fundamental's energy above 1.3 eV, which can be seen more clearly in the calculated $\phi_{MoW}$ shown in Fig. 3d. We also note that the theory predicted the experimental data reasonably well. Viewing the amplitude of $\chi^{(2)}$ dictating SHG intensities (Fig. S7d) and optical absorption,[29] the rise of $\phi_{MoW}$ at 1.3 eV was attributed to the distinctive band structures and unequal optical transitions mostly by C excitons at ~2.8 eV (= $2\hbar\omega$) in the two materials.[5, 29] The energy region above 1.6 eV where $\phi_{MoW}$ increased further is occupied with intense optical transitions of D excitons.[30] In the picture of a driven harmonic oscillator, finite damping (light absorption) at second-harmonic frequency leads to a phase delay with respect to the fundamental driving field.[17] Then nonzero $\phi_{MoW}$ and its frequency dependence are due to material-dependent resonance frequencies and damping.

Our findings not only bear the stated implications for the emerging nonlinear optics of 2D materials but will also lead to important practical applications. VdW stacks of 2D crystals can be an excellent miniature photonic system that may allow custom-designed parametric generation. The demonstration of electrical tuning of SHG intensity[31] suggests that phase modulation is also feasible in these atom-thick nonlinear optical materials. Our approach also enables the optical determination of stack angles. As can be readily seen in Fig. 1f, the min/max intensity ratio is directly related to the stack angle of a given sample. This method does not require probing unstacked 1L areas and can be extended to multi-stacks beyond bilayers, as the phase information completes the description of SHG processes. Additionally, the phase resolution endowed by the interferometry will allow complete differentiation of crystallographic domains, which was limited in the conventional intensity-based polarized SHG imaging.[22]

In summary, we reported the optical second-harmonic interference occurring in the two-dimensional limit of atom thickness. The SHG signals from artificial 2D heterocrystals of $MoS_2/WS_2$ underwent coherent superposition and exhibited complicated polarization behavior for varying stack angle and photon energy. Using spectral phase interferometry and polarized SHG, we directly measured the inter-material difference of the SHG phase originating from differential interactions of both materials with light. First-principles calculations on second-order susceptibilities revealed its electronic origins also verifying the superposition model. This work will also contribute toward creating novel nonlinear optical and photonic applications using low-dimensional materials.

**Methods**

**Preparation of samples.** Single-layer $WS_2$ and $MoS_2$ samples were prepared by mechanical exfoliation[32] of commercial bulk crystals (2D semiconductors). To avoid unwanted optical interference, amorphous quartz was used as substrates after thorough cleaning using piranha solutions. Homo and hetero-bilayers were prepared by transferring a top layer exfoliated on a polydimethylsiloxane (PDMS) film onto a bottom layer supported on a quartz substrate and substrate was heated to a 60 °C to enhance interlayer coupling during the process.[27] For samples whose interlayer distance is larger than 2 nm, post vacuum annealing was carried out at 200 °C for 3 h. For a targeted stack angle, the crystallographic orientation of each layer was first determined by polarized SHG measurements. Then the transfer was made with the angular alignment of the top layer with respect to the bottom one. The positional and angular errors for the targeted transfer were less than 2 microns and 1 degree, respectively.

**SHG measurements.** SHG measurements were performed with a home-built micro-SHG spectroscopy setup configured upon a commercial microscope (Nikon, Ti-U) as described elsewhere.[33] As a fundamental pulse, the plane-polarized output from a tunable Ti:sapphire laser (Coherent Inc., Chameleon) was focused on samples using a microscope objective (40X, numerical aperture = 0.60). The FWHM of the focal spot was 2.3 ± 0.2 μm. The pulse duration and repetition rate were 140 fs and 80 MHz, respectively. The backscattered SHG signals were collected with the same objective and guided to a spectrometer equipped with a thermoelectrically cooled CCD detector (Andor Inc., DU971P). To vary the direction of the fundamental's polarization with respect to exfoliated 2D crystals, they could be rotated about a surface normal to their basal planes using a rotational mount. For polarization-resolved measurements, an analyzing polarizer was placed in front of the spectrometer to select polarization components of interest. Samples could be rotated with a precision of 0.2 degree using a rotational mount.

For the spectral phase interferometry (Supplementary Section S3), reference SHG pulses were generated in an in-line geometry (Fig. 3a) by focusing the fundamental beam at a z-cut α-quartz window of 100-μm thickness. The time delay between the SHG pulse by samples and the reference SHG pulse was maintained in the range of 1 ~ 3 ps by inserting dispersing materials including N-BK7 glass windows. Because the refractive objective induced excess time delay, it was replaced with a Cassegrain-type reflective lens (Edmund optics, 52X, numerical aperture = 0.65). To prevent sample degradation, the average power density of the incident fundamental beam was maintained below 40 μW/μm$^2$ for the interferometric measurements and 100 μW/μm$^2$ for the others.

**First-principles calculations.** For density functional theory (DFT) calculation of nonlinear optics, the ELK package with all-electron augmented plane wave methods were used under the generalized gradient approximation of Perdew–Berke–Ernzerhof (GGA-PBE) [34, 35]. For the accurate calculation of second-order susceptibility tensor $\chi^{(2)}$,[36] we used dense 72×72×1 k-mesh also considering the effects of spin-orbit coupling. The expression for the transition probability was derived from the second-order perturbation theory[36] (Supplementary Section S4). The functional form of $\chi^{(2)}$ of monolayer TMD is given in Supplementary Section S1. As DFT calculation typically underestimates electronic bandgaps, scissors shifts were applied to compensate for the discrepancy.[37] Scissors shifts of 0.055 and 0.11 eV respectively for $MoS_2$ and $WS_2$ monolayers were found to reproduce well the experimental phase difference between the two materials.

**Supporting information**

Effects of thermal annealing on morphology of bilayers, Raman and photoluminescence spectra of $2L_{MoW}$, SHG polar graphs and spectra of $2L_{MoW}$ samples with varying stack angles, insensitivity of SHG to thermal annealing, validation of SHG spectral interferometry, interferometric measurements and raw data of $2L_{MoW}$, second-order susceptibility of $1L_{Mo}$ and $1L_W$, scheme for angle-resolved SHG, polarization dependence of SHG from 1L $MX_2$ belonging to $D_{3h}^1$ space group, second-harmonic interference in hetero and homo-bilayers, direct determination of phase difference in hetero-bilayers using spectral phase interferometry, calculation of complex second-order susceptibility.

**Author contributions:** S.R. proposed and directed the research; W.K. and J.O. carried out experiments and data analysis; J.S. and J.A. performed theoretical simulations and analysis; All discussed the results and contributed in writing the manuscript.

**Competing interests:** We declare no competing interests.

**Acknowledgements**

S.R. acknowledges the financial support from the National Research Foundation of Korea (NRF-2016R1A2B3010390, 2019R1H1A2079871, and NRF-2020R1A2C2004865). J.H.S. was financially supported by the National Research Foundation of Korea (NRF-2020R1A5A1019141).

**References**

1. Bonaccorso, F.; Sun, Z.; Hasan, T.; Ferrari, A. C., Graphene photonics and optoelectronics. *Nat. Photonics* **2010,** *4* (9), 611-622.
2. Schaibley, J. R.; Yu, H.; Clark, G.; Rivera, P.; Ross, J. S.; Seyler, K. L.; Yao, W.; Xu, X., Valleytronics in 2D materials. *Nat. Rev. Mater.* **2016,** *1* (11), 16055.
3. Tran, T. T.; Bray, K.; Ford, M. J.; Toth, M.; Aharonovich, I., Quantum emission from hexagonal boron nitride monolayers. *Nat. Nanotechnol.* **2016,** *11* (1), 37-41.
4. Chernikov, A.; Berkelbach, T. C.; Hill, H. M.; Rigosi, A.; Li, Y.; Aslan, O. B.; Reichman, D. R.; Hybertsen, M. S.; Heinz, T. F., Exciton Binding Energy and Nonhydrogenic Rydberg Series in Monolayer WS2. *Phys. Rev. Lett.* **2014,** *113* (7), 076802.
5. Malard, L. M.; Alencar, T. V.; Barboza, A. P. M.; Mak, K. F.; de Paula, A. M., Observation of intense second harmonic generation from MoS2 atomic crystals. *Phys. Rev. B* **2013,** *87* (20), 201401.
6. Wang, G.; Marie, X.; Gerber, I.; Amand, T.; Lagarde, D.; Bouet, L.; Vidal, M.; Balocchi, A.; Urbaszek, B., Giant Enhancement of the Optical Second-Harmonic Emission of WSe2 Monolayers by Laser Excitation at Exciton Resonances. *Phys. Rev. Lett.* **2015,** *114* (9), 097403.
7. Li, Y.; Rao, Y.; Mak, K. F.; You, Y.; Wang, S.; Dean, C. R.; Heinz, T. F., Probing Symmetry Properties of Few-Layer MoS2 and h-BN by Optical Second-Harmonic Generation. *Nano Lett.* **2013,** *13* (7), 3329-3333.
8. Zhou, X.; Cheng, J.; Zhou, Y.; Cao, T.; Hong, H.; Liao, Z.; Wu, S.; Peng, H.; Liu, K.; Yu, D., Strong Second-Harmonic Generation in Atomic Layered GaSe. *J. Am. Chem. Soc.* **2015,** *137* (25), 7994-7997.
9. Franken, P. A.; Hill, A. E.; Peters, C. W.; Weinreich, G., Generation of Optical Harmonics. *Phys. Rev. Lett.* **1961,** *7* (4), 118-119.
10. Fan, T. Y.; Byer, R. L., Diode laser-pumped solid-state lasers. *IEEE J. Quantum Electron.* **1988,** *24* (6), 895-912.
11. Corn, R. M.; Higgins, D. A., Optical second harmonic generation as a probe of surface chemistry. *Chem. Rev.* **1994,** *94* (1), 107-125.
12. Campagnola, P., Second Harmonic Generation Imaging Microscopy: Applications to Diseases Diagnostics. *Anal. Chem.* **2011,** *83* (9), 3224-3231.
13. Wen, X.; Gong, Z.; Li, D., Nonlinear optics of two-dimensional transition metal dichalcogenides. *InfoMat* **2019,** *1* (3), 317-337.
14. Wang, Y.; Xiao, J.; Yang, S.; Wang, Y.; Zhang, X., Second harmonic generation spectroscopy on two-dimensional materials. *Opt. Mater. Express* **2019,** *9* (3), 1136-1149.
15. Zeng, H.; Cui, X., An optical spectroscopic study on two-dimensional group-VI transition metal dichalcogenides. *Chem. Soc. Rev.* **2015,** *44* (9), 2629-2642.
16. Novoselov, K. S.; Mishchenko, A.; Carvalho, A.; Castro Neto, A. H., 2D materials and van der Waals heterostructures. *Science* **2016,** *353* (6298), aac9439.
17. Chang, R. K.; Ducuing, J.; Bloembergen, N., Relative Phase Measurement Between Fundamental and Second-Harmonic Light. *Phys. Rev. Lett.* **1965,** *15* (1), 6-8.
18. Kemnitz, K.; Bhattacharyya, K.; Hicks, J. M.; Pinto, G. R.; Eisenthal, B.; Heinz, T. F., The phase of second-harmonic light generated at an interface and its relation to absolute molecular orientation. *Chem. Phys. Lett.* **1986,** *131* (4), 285-290.
19. Veenstra, K. J.; Petukhov, A. V.; de Boer, A. P.; Rasing, T., Phase-sensitive detection technique for surface nonlinear optics. *Phys. Rev. B* **1998,** *58* (24), R16020-R16023.
20. Yazdanfar, S.; Laiho, L. H.; So, P. T. C., Interferometric second harmonic generation microscopy. *Opt. Express* **2004,** *12* (12), 2739-2745.

21. Hsu, W.-T.; Zhao, Z.-A.; Li, L.-J.; Chen, C.-H.; Chiu, M.-H.; Chang, P.-S.; Chou, Y.-C.; Chang, W.-H., Second Harmonic Generation from Artificially Stacked Transition Metal Dichalcogenide Twisted Bilayers. *ACS Nano* **2014**, *8* (3), 2951-2958.
22. Yin, X.; Ye, Z.; Chenet, D. A.; Ye, Y.; O'Brien, K.; Hone, J. C.; Zhang, X., Edge Nonlinear Optics on a MoS2 Atomic Monolayer. *Science* **2014**, *344* (6183), 488-490.
23. Schaibley, J. R.; Rivera, P.; Yu, H.; Seyler, K. L.; Yan, J.; Mandrus, D. G.; Taniguchi, T.; Watanabe, K.; Yao, W.; Xu, X., Directional interlayer spin-valley transfer in two-dimensional heterostructures. *Nat. Commun.* **2016**, *7* (1), 13747.
24. Hu, D.; Chen, K.; Chen, X.; Guo, X.; Liu, M.; Dai, Q., Tunable Modal Birefringence in a Low-Loss Van Der Waals Waveguide. *Adv. Mater.* **2019**, *31* (27), 1807788.
25. Liu, C.-H.; Clark, G.; Fryett, T.; Wu, S.; Zheng, J.; Hatami, F.; Xu, X.; Majumdar, A., Nanocavity Integrated van der Waals Heterostructure Light-Emitting Tunneling Diode. *Nano Lett.* **2017**, *17* (1), 200-205.
26. Castellanos-Gomez, A.; Buscema, M.; Molenaar, R.; Singh, V.; Janssen, L.; van der Zant, H. S. J.; Steele, G. A., Deterministic transfer of two-dimensional materials by all-dry viscoelastic stamping. *2D Mater.* **2014**, *1* (1), 011002.
27. Ryu, Y.; Kim, W.; Koo, S.; Kang, H.; Watanabe, K.; Taniguchi, T.; Ryu, S., Interface-Confined Doubly Anisotropic Oxidation of Two-Dimensional MoS2. *Nano Lett.* **2017**, *17* (12), 7267-7273.
28. Lee, C.; Yan, H.; Brus, L. E.; Heinz, T. F.; Hone, J.; Ryu, S., Anomalous Lattice Vibrations of Single- and Few-Layer MoS2. *ACS Nano* **2010**, *4* (5), 2695-2700.
29. Li, Y.; Chernikov, A.; Zhang, X.; Rigosi, A.; Hill, H. M.; van der Zande, A. M.; Chenet, D. A.; Shih, E.-M.; Hone, J.; Heinz, T. F., Measurement of the optical dielectric function of monolayer transition-metal dichalcogenides: MoS2, MoSe2, WS2, and WSe2. *Phys. Rev. B* **2014**, *90* (20), 205422.
30. Frindt, R. F.; Yoffe, A. D., Physical properties of layer structures : optical properties and photoconductivity of thin crystals of molybdenum disulphide. *Proc. R. Soc. A-Math. Phys. Eng. Sci.* **1963**, *273* (1352), 69.
31. Seyler, K. L.; Schaibley, J. R.; Gong, P.; Rivera, P.; Jones, A. M.; Wu, S.; Yan, J.; Mandrus, D. G.; Yao, W.; Xu, X., Electrical control of second-harmonic generation in a WSe2 monolayer transistor. *Nat. Nanotechnol.* **2015**, *10* (5), 407-411.
32. Novoselov, K. S.; Geim, A. K.; Morozov, S. V.; Jiang, D.; Zhang, Y.; Dubonos, S. V.; Grigorieva, I. V.; Firsov, A. A., Electric field effect in atomically thin carbon films. *Science* **2004**, *306* (5696), 666-669.
33. Ko, T. Y.; Jeong, A.; Kim, W.; Lee, J.; Kim, Y.; Lee, J. E.; Ryu, G. H.; Park, K.; Kim, D.; Lee, Z.; Lee, M. H.; Lee, C.; Ryu, S., On-stack two-dimensional conversion of MoS2 into MoO3. *2D Mater.* **2017**, *4* (1), 014003.
34. Dewhurst, K.; Sharma, S.; Nordström, L.; Cricchio, F.; Bultmark, F.; Granas, O.; Gross, H. The ELK code. http: //elk.sourceforge.net/ (Feb. 20, 2020).
35. Sharma, S.; Ambrosch-Draxl, C., Second-Harmonic Optical Response from First Principles. *Phys. Scr.* **2004**, *T109*, 128.
36. Sipe, J. E.; Moss, D. J.; van Driel, H. M., Phenomenological theory of optical second- and third-harmonic generation from cubic centrosymmetric crystals. *Phys. Rev. B* **1987**, *35* (3), 1129-1141.
37. Levine, Z. H.; Allan, D. C., Linear optical response in silicon and germanium including self-energy effects. *Phys. Rev. Lett.* **1989**, *63* (16), 1719-1722.

**TOC Figure**

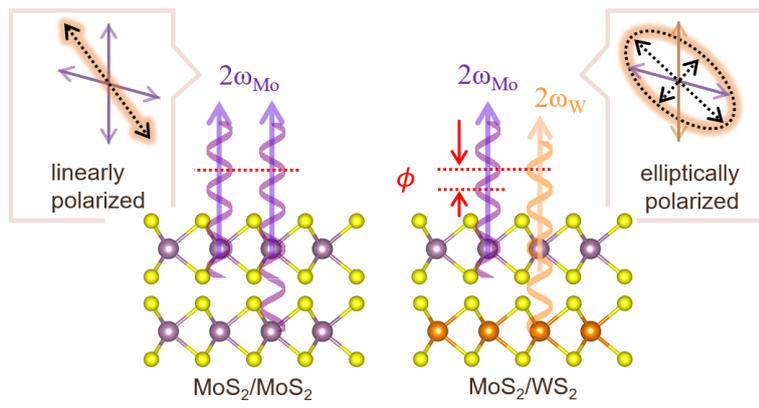